\documentclass[11pt,twoside]{article}

\usepackage{asp2006}
\usepackage{epsf}
\usepackage{psfig}
\usepackage{lscape}
\usepackage{color}

\newcommand{\Msun}{M$_\odot$}

\begin{document} 

\title{A Bayesian Approach Accounting for Stochastic Fluctuations in Stellar
Cluster Properties}   
\author{M. Fouesneau \& A. Lan\c con}   
\affil{Observatoire Astronomique(UMR7550), Universit\'e de Strasbourg \& CNRS,
11 rue de l'Universit\'e, 67000 Strasbourg, France}    

\begin{abstract} 
The integrated spectro-photometric properties of star clusters are subject to 
large cluster-to-cluster variations. They are distributed in non trivial ways around 
the average properties predicted by standard population synthesis models. 
This results from the stochastic mass distribution of the finite (small) 
number of luminous stars in each cluster, stars which may be either particularly
blue or particularly red. The color distributions are broad and usually far from
Gaussian, especially for young and intermediate age clusters, as found in
interacting galaxies. When photometric measurements of clusters are used
to estimate ages and masses in conjunction with standard models,
biases are to be expected.  We present a Bayesian approach that explicitly 
accounts for stochasticity when estimating ages and
masses of star clusters that cannot be resolved into stars. Based on
Monte-Carlo simulations, we are starting to explore the probability distributions 
of star cluster properties obtained given a set of multi-wavelength photometric data. 
\end{abstract}

\section{Introduction} 
Observing star clusters offers a powerful tool for studies of the formation
history of galaxies. Clusters can be measured out to much further distances 
than individual field stars, and they provide complementary information to the 
integrated light emitted by galaxy fields.
Star formation and cluster formation are closely related, although the
interplays between clusters and the field remain to be properly understood.
The age distribution of star clusters may be a record of the star formation 
history of galaxies. Their age and mass distributions are especially important 
when we wish to understand what happens during collisions or mergers of 
gas-rich galaxies, where the interaction is well known to trigger 
the formation of stars.

For clusters that are not resolved into stars, estimates of ages, masses and other 
intrinsic properties are based on integrated photometry or spectra.
The standard approach is to compare
these observed properties with synthetic spectra produced by population synthesis codes
that assume a continuously populated stellar initial mass function (IMF).

\begin{figure}
        \plotone{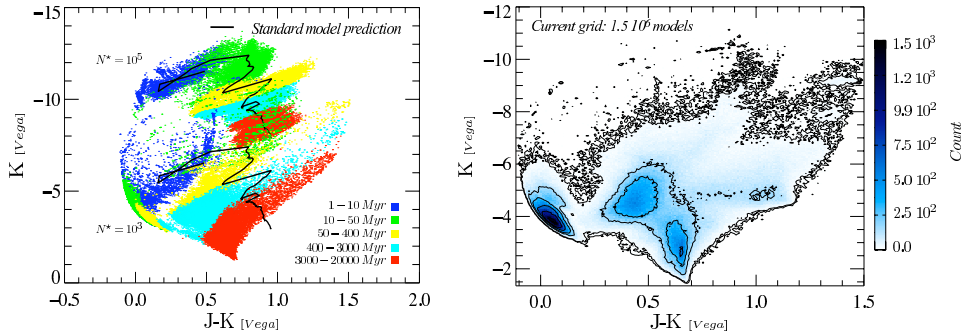}
        \caption{Stochastic properties of star clusters
	at solar metallicity. 
	\textbf{Left}: The dots represent photometric properties of 
	individual clusters containing $10^3$ and $10^5$ stars each.
        The solid lines show the corresponding age sequences from 
	``standard" predictions, i.e. rescaled versions of population synthesis 
	predictions strictly valid only for a very large number of stars 
	(figure inspired by \citealt{Bruzual2002}).
	\textbf{Right}: Density distribution of models in our catalog,
        constructed assuming a power law with a index -2 for the cluster
	mass function. 
        \label{fig:IRspread}
        }
\end{figure}

However the energy distribution of clusters is dominated by the light of a
finite (often small) number of massive stars, 
whereas the mass is mostly produced by 
early evolved stars with relatively faint luminous contributions. 
As an example, red supergiants are the predominant source of near-IR light
in young clusters (6-60\,Myr). These stars are
intrinsically rare. As a combined result of the Initial Mass Function
(IMF) and the short duration of the relevant evolutionary phases, 
we expect on average only one red supergiant for a total cluster mass 
of $10^4$ \Msun~for a standard IMF with a
lower cut-off of $0.1$ \Msun \citep{Lancon2008}. Hence a $10^4$ \Msun~star
cluster may contain zero, one or a few red supergiants, which will result
in multi-modal probability distributions for integrated colors and fluxes 
\citep{Chiosi1988, Lancon2000, Cervino2004,Lancon2009}. 
Large spreads  result from the above-mentioned small numbers of 
luminous stars, the so-called stochastic fluctuations (Figure \ref{fig:IRspread}). 
It requires clusters with several $10^6$ \Msun~to narrow down 
the fluctuations to 5\% in the K-band. In this case, colors and fluxes follow
uni-modal distributions and Gaussian approximations become tolerable.

Massive clusters for which the Gaussian approximation  holds
exist in very small numbers, because cluster mass functions 
tend to fall rapidly with increasing mass \citep{Whitmore1999, Bastian2009}. 
On the other hand, many low and 
intermediate mass clusters have already been observed with space-based and
large ground-based telescopes. It is thus very important to develop
methods that explicitely deal with the stochastic fluctuations.

\section{Estimates of Star Cluster Properties}
The method we are developing follows a Bayesian approach. 
It is a close analog to the one introduced by
\citet{Kauffmann2003} for the study of star formation histories in the Sloan
Digital Sky Survey. However the variety of observable properties has
completely different origins in both contexts: stochasticity at a given
age, mass and metallicity, plays a predominant role here while it is 
negligible in most galaxies taken as a whole.

We have started a campaign of Monte-Carlo (MC) simulations 
(Fig.\,\ref{fig:IRspread}), that is
used as a catalog for establishing the joint probability distributions of
the intrinsic parameters of a cluster (age, total number of stars or total 
mass, metallicity, extinction), given a set
of photometric measurements and their uncertainties (Fouesneau et al., in prep).
The total mass is not a simple scaling factor. It
enters the problem even when only colors are available 
because it affects the color probability distributions of clusters of 
a given age and metallicity.
 
Among the assumptions in the method are the stellar IMF, and the 
adopted stellar evolutionary tracks and stellar spectral libraries.
The main {\em a priori} of the Bayesian inversion are the underlying 
cluster formation history (number of clusters per age bin in the 
catalog), and the cluster mass function. In applications to real
cluster populations, several {\em a priori} distributions should be 
tested and compared. The effects of dynamical evolution on the IMF
and on the cluster mass function also need to be discussed.

\begin{figure}
        \plotone{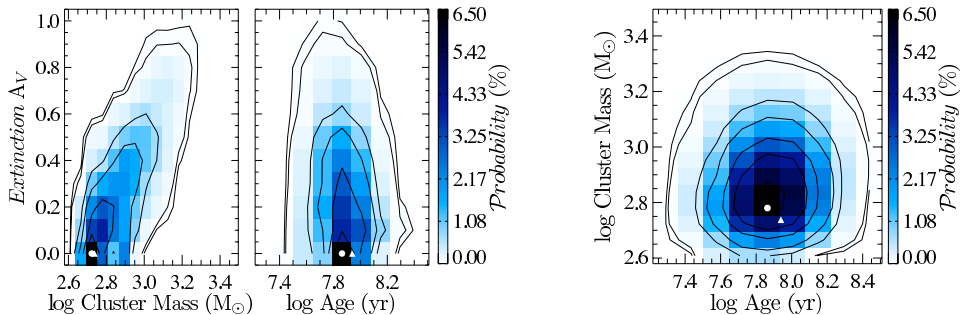}
        \caption{
	Example of {\em a posteriori} probability distributions of age, mass
 and extinction, for a synthetic cluster ``observed" in U, B, V, I, J and K 
 with 1-$\sigma$ photometric errors of 0.05 magnitudes. 
 The prior is that the cluster mass distributions in galaxies vary as 
 $M^{-2}$, and that the cluster age distributions in galaxies are
 constant in logarithmic age units (only clusters that survived dynamical
 disruption matter here).  The true properties of the ``observed" 
 cluster are indicated as triangles, the most probable values  as
 derived from the ``observations" are shown as small circles.
        \label{fig:SEDex}
        }
\end{figure}

We generate integrated photometric properties of individual clusters with
a modified version of P\'EGASE \citep{Fioc1999} that implements a discrete
IMF. 
The models take into account the nebular continuum and line emission of 
clusters that contain ionizing stars.

The assumption that errors in the observational data are
Gaussian leads to a simple expression for the {\em a posteriori}
probability distribution of an intrinsic property, 
such as the age or mass of an individual cluster. 
The probability for
an intrinsic property $X$ to be located in an interval $[x_1,x_2]$, given 
the photometric measurements $Y=\{Y_k\}_{k\in[1..n]}$, is:
\begin{eqnarray}
	\mathcal{P}\,(\ X\ \in\ [x_1,x_2]\ |\ Y\ )\  =\ \alpha \times 
		\qquad \qquad \qquad \qquad \qquad \qquad \qquad\nonumber \\
	\qquad \sum_{i,\,X(M_i)\in[x_1,x_2]} \ \prod_k \frac{1}{\sqrt{2\pi\sigma_k^2}}
	\ \exp{
	\left( -\frac{\left( Y_k - Y_{k,M_i} \right)^2}{2\sigma_k^2} \right)}
	\times \mathcal{P}(M_i),
\label{eq:proba}
\end{eqnarray}
where $\mathcal{P}(M_i)$ is the probability of the individual model $M_i$ to be chosen
and $\alpha$ is the normalization constant.
Through the factors $\mathcal{P}(M_i)$, this expression adapts to any prior 
mass or age distribution.

As an example, Figure \ref{fig:SEDex} shows recovered age, 
mass and extinction probability distributions based on pseudo-observational
data, i.e. on the synthetic magnitudes of one of the clusters in the MC-model
catalog (modified by the addition of reasonable noise). Although  
six absolute fluxes across the spectrum are used, the possible 
age and mass ranges remain significantly broad, and asymmetric.
We are in the process of establishing the systematic differences between 
the actual properties of a cluster and various estimates thereof (best $\chi^2$ model,
most probable values, and values recovered using the classical
approach with a continuously populated IMF). We will also reinvestigate
which minimal combination of observations constrains the
intrinsic properties most efficiently.

Varying the metallicity, the allowed extinction range or the 
{\em a priori} distributions of mass and age will significantly change the 
results. We are still exploring the different effects, 
especially those of metallicity and extinction. We are also 
extending the MC-model catalog by adding new massive clusters,
which will be necessary to test the effect of adopting flatter
{\em a priori} cluster mass distributions. 

\section{Conclusion}
Although massive clusters contain important information on
active episodes of star formation in galaxies, the need to
study smaller clusters is increasing as the amount of available 
observational data grows.  Intrinsically rare but luminous stars 
are responsible for large stochastic fluctuations in the integrated energy distribution 
of these objects.  Hence, new tools that deal with stochastic fluctuations explicitly need to be 
developed.  In this context we have implemented a Bayesian approach that, given a 
set of photometric data, provides most probable cluster parameters as well as
confidence intervals. The ongoing study of systematic differences between true and 
estimated values should allow us to correct any significant biases in previous studies.

\acknowledgements We thank M. Fioc for making his updated version of {\sc P\'egase}
available for modification in advance of publication.


\end{document}